\documentstyle[11pt]{article}
\def\rmd{{\rm d}}
\def\etal{{\it et al.}}
\def\half{{\textstyle{\frac{1}{2}}}}
\def\plab{\mbox{{$p$\raisebox{-0.25em}{{\small{\sl lab}}}}}}
\def\Asym{\mbox{{$A$\raisebox{-0.25em}{{\small{\small{\sl 0}}{\sl n}}}}}}
\def\Del{{\mit \Delta}}

\def\GeVc{{\rm GeV\hspace{-0.5mm}/\hspace{-0.5mm}c}}
\def\dsdo{{\rmd \sigma\!/\!\rmd {\mit \Omega}}}
\newcommand{\pbar}{\overline{p}}
\newcommand{\pbarp}{\overline{p}p}
\newcommand{\pbarppipi}{\overline{p} p  \rightarrow \pi^- \pi^+}
\newcommand{\pbarpKK}{\overline{p}p \rightarrow K^- K^+}
\newcommand{\qbarq}{\overline{q}q}
\newcommand{\NbarN}{\overline{N} N}
\newcommand{\KbarK}{\overline{K} K}
\newcommand{\pipi}{\pi \pi}

\newcommand{\simg}{\parbox{0.9em}{\raisebox{0.3ex}
	{$>$}\hspace{-0.8em}\raisebox{-0.3em}{$\sim$}}}
\newcommand{\siml}{\parbox{0.8em}{\raisebox{0.3ex}
	{$<$}\hspace{-0.8em}\raisebox{-0.3em}{$\sim$}}}
\begin{document}

\begin{center}

\vspace*{15mm}

\title
{On the Spectacular Large Asymmetry
  in $\pbarppipi$ and $\pbarpKK$ Reactions}

\vspace{5mm}

\author{
S. Takeuchi,
F. Myhrer and K. Kubodera \\
Department of Physics and Astronomy,
 University of South Carolina, \\
Columbia, SC29208 }

\date{}

\maketitle

\vspace{5cm}

Talk given at the Workshop of Future Directions in Particle and
Nuclear Physics at Multi-GeV Hadron Facilities, \\
Brookhaven Nat. Lab., March, 1993.

\end{center}

\newpage

\begin{center}


\title
{On the Spectacular Large Asymmetry
  in $\pbarppipi$ and $\pbarpKK$ Reactions}

\vspace{5mm}

\author{
S. Takeuchi,
F. Myhrer and K. Kubodera \\
Department of Physics and Astronomy,
 University of South Carolina, \\
Columbia, SC29208 }

\date{}

\maketitle

{\bf Abstract.}

\end{center}

An illustrative analysis is presented to show
the origin of the energy-independent
maximal asymmetry observed for wide ranges of angles
in the reactions $\pbarppipi$ and $\pbarpKK$.
The general nature of our simple relation
between helicity  -flip and -nonflip partial wave amplitudes
enforces the notion that
these features of the asymmetry
for these two
annihilation reactions
are likely to persist within the hadronic regime.
At higher energies these features of the asymmetry
will probably be modified significantly,
signaling the onset of perturbative QCD.
Our study supports the arguments that the final $\KbarK$ state
originates from a more central reaction than the $\pipi$
final state.


\section{ Introduction}
The experimentally observed asymmetries \Asym\ in
the annihilation reactions $\pbarppipi$ and $\pbarpKK$
seem to reach the maximal possible
value of 1 over wide ranges of angles and \plab\
between about 1 GeV/c and 2.2 GeV/c \cite{Ei75,Te89,Hasan92}.
The reaction $\pbarppipi$ has a
very large \Asym\ for \plab\ \simg 1.5 \GeVc, while
the \Asym\ of the final $K^-K^+$ state has values
close to 1
for \plab\ \simg 1 \GeVc.
These remarkable features seem
to call for a simple explanation.
This explanation must simultaneously account for the
following aspects of the observed differential cross sections:
the $\dsdo$ for the final $\pi^-\pi^+$ reaction
shows pronounced oscillations
whereas that of the final $K^-K^+$
reaction
has a strong forward peak and a smooth backward plateau.

The angular oscillations of the
$\dsdo$ for $\pbarppipi$
lead in an early model analysis
to the speculation of the existence of
possible J=3, 4 and 5 meson resonances \cite{Ca77},
where it was assumed
that one partial wave dominates at each energy.
A recent partial wave analysis based on dispersion relation theory
of the $\pbarppipi$ reaction is not incompatible with
"resonance activity" in some partial waves \cite{MO88}.
Our "geometrical" analysis \cite{TMK} presented below
does not require any explicit meson resonances and will
reproduce in a natural way the observed behaviour of $\Asym$
and $\dsdo$.

\medskip

{}From the baryon-meson picture
these reactions with two pseudo-scalar mesons in the final state
are expected to be similar to each other.
{}From a subnucleonic viewpoint
they can be very different in nature
because  $\pbarpKK$ involves
the annihilation of two initial valence $\qbarq$-pairs
accompanied by the creation of an $\overline{s}s$-pair,
while $\pbarppipi$ can take place
simply by annihilating one $\qbarq$-pair.
Which picture is more appropriate may depend on the energy.
The angular dependence of the measured  \Asym\ and
$\dsdo$ for the two reactions
\cite{Ei75,Te89} indicates that
the ``reaction mechanism" for \plab\ \siml 1 \GeVc\
is different from that of the higher energies.
At these low energies
the coupled-channels method with the explicit enumeration of possible
hadronic channels
may be a useful theoretical framework \cite{LT90,MHS91}.
As the incident energy increases, the coupled-channels method becomes
more and more complicated;
meanwhile, in the few GeV/c energy region,
it is expected that we are still below the energy regime
where perturbative QCD calculations are valid.
The afore-mentioned maximum symmetry is seen
at the higher end of the LEAR energies,
and our studies so far \cite{TMK} have focused on
this energy region.
At the $AGS$ accelerator at Brookhaven National Laboratory
and at the proposed $SuperLEAR$ and $KAON$
facilities we probably will reach
energies where perturbative QCD calculations become
relevant for exclusive hadronic reactions.
For these higher energies,
we expect that the observed maximum asymmetry phenomena
{\it will break down}, signaling the onset of the
perturbative QCD energy regime.

\section{ A Diffraction Model Analysis}

Each of the reactions under consideration can be
characterized by two independent helicity amplitudes:
$f_{++}$ (helicity non-flip) and $f_{+-}$ (helicity flip).
In terms of
these two amplitudes
the cross section and the asymmetry are given as
\vspace{-2mm}
\begin{equation}
   \dsdo =  | f_{++} |^2 +  |f_{+-} |^2
\hspace{0.8cm}{\rm and}\hspace{0.8cm}
\Asym = 2 \Im m(f_{++}^* f_{+-}) / (\dsdo).
\label{b}
\vspace{-2mm}
\end{equation}
The partial wave expansion gives
\vspace{-2mm}
\begin{equation}
 f_{++} = \frac{1}{p} \sum_{J=0}^{\infty} \: (J + \half) \:
T_{+}^{J} \: P_{J}(\cos \theta)
\label{c}
\vspace{-2mm}
\end{equation}
and
\vspace{-2mm}
\begin{equation}
 f_{+-} = \frac{1}{p} \sum_{J=0}^{\infty} \:
 (J+ \half )/ \sqrt{J(J+1)} \;\; T_{-}^J
 \: P_{J}^{\prime}(\cos \theta) \: \sin \theta .
\label{d}
\vspace{-1mm}
\end{equation}
Conservation of parity and angular momentum implies that
only tensor-coupled
$\NbarN$ partial waves ($J = L \pm 1$) contribute to this reaction.
We assume  that $T_-^J$ is given by the derivative
of $T_+^J$ w.r.t the
scattering impact parameter $b$ since we expect
the helicity flip amplitude is most effective at the
{\it interaction surface:}
\vspace{-2mm}
\begin{equation}
   T_-^J = {\rm const.} \; \partial T_+^J / \partial b .
\label{f}
\vspace{-2mm}
\end{equation}
(A similar relation
has been found phenomenologically
in the corresponding $t$-channel process, $\pi N$ scattering \cite{Ho71},
as well as in $\pbarp$ elastic scattering \cite{Ei76} in
this momentum range.)
Then, using $J \approx pb$, we find the basic
``differential" relation,
\vspace{-2mm}
\begin{equation}
   T_{-}^J \; \propto\; \Del T_{+}^J / \Del J
\end{equation}
or
\begin{equation}
%
   \frac{J + \half}{\sqrt{J(J + 1)}}\; T_{-}^J = \frac{1}{\beta}\:
 (T_{+}^{J-1} - T_{+}^{J+1}) .
\label{main}
\vspace{-2mm}
\end{equation}
where $\beta$ is a constant parameter. This assumption leads to
\begin{equation}
    f_{+-} = - \frac{1}{\beta} \: f_{++} \sin \theta
\label{main2}
\vspace{-2mm}
\end{equation}
and
\begin{equation}
\Asym = \frac{2 \Im m \beta}{| \beta|^2 + \sin^2 \theta} \: \sin \theta
\label{asym}
\vspace{-2mm}
\end{equation}
With an imaginary $\beta (=i)$, \Asym\ of eq.(\ref{asym}) will be
larger than 0.9 over a very wide angular range
($| \cos \theta | \sim$ 0.8) whereas $\dsdo$ may have a
significantly stronger angular dependence, determined by
$f_{++}(\theta)$. (We can show in a DWBA- type calculation that
$\beta$ is almost constant as a function of $\theta$ \cite{TMK}.)

\medskip

Since so many competing annihilation channels
 are open at the energies under discussion, we initially
assume as an explicit model example that the amplitudes
are given by ``classical" grey- or black-sphere amplitudes.
These amplitudes will give
\Asym\ $\approx$ 1.
%
%
\begin{equation}
   T_{+}^J = \left\{\begin{array}{ll}  B \ exp(-aJ) & (J \leq J_{max}) \\
               0 &  (J \geq J_{max}) \\ \end{array} \right.
\end{equation}
where $B$ and $a$ are constants. To reproduce the observed $\dsdo$ for
$\pbarppipi$ we need $a \approx$ 0 and $J_{max}$ = 4.
This is a "black" sphere amplitude of radius equal to $J_{max}/q$.
For the reaction $\pbarpKK$ data requires
$a \approx$ 0.5 corresponding to
a "grey" sphere.
This means the  lowest partial waves (J=0 and 1) dominate in
this reaction.
As a consequence we conclude that the $\pbarppipi$ reaction occurs over
a larger interaction volume than the $\pbarpKK$ reaction \cite{TMK}.

\section{Discussion and Conclusions}

We have also studied this problem using a DWBA approach \cite{TMK}
to further examine the assumption of the phenomenological analysis
above.
The optical potential for the initial $\NbarN$ channel should reflect
the strong $\NbarN$ absorption for the low impact parameter region.
This implies that
the low partial wave amplitudes are close to their unitarity limit.
We also expect strong absorption effects
from the final-state $\pi \pi$ and $\KbarK$
interactions.
However, because the final state interaction is not well known
at these high energies,
we parametrize the effective transition operator that simulates the
combined effects of the final state interaction and the transition operator.

\medskip

Our study \cite{TMK} indicates:
\begin{enumerate}
\item
The strong angular dependent cross section  and the
smooth angular asymmetry can be reproduced simultaneously
with the use of a simple transition operator potential.
This potential must be a sum of two terms of very different ranges.
\item
The interference term arising from
the sum of the two transition potentials
(interference of the final state interaction and the annihilation
reaction?)
is essential to yield the large asymmetries.
This is consistent with the findings at low energies (\plab\ \siml 1 \GeVc)
 \cite{MHS91};
\item
The resulting behaviour of the "effective" $\beta$ (see
eq.(\ref{asym}) of the diffraction model) is found to be almost
independent of angle $\theta$ and dominantly imaginary.
\item
The initial state distortion due to our $\NbarN$ $L \cdot S$
potential plays a minor role in explaining \Asym.
(Since it is tensor coupled $\NbarN$ partial waves which contribute
to these reactions we should have used $\NbarN$ tensor forces to
generate the initial $L \cdot S$ $\NbarN$ amplitudes.
\item
As stated the $\pbarpKK$ reaction takes place at much shorter
distances than the $\pbarppipi$ reaction. This result lends support
to the arguments based on an analogy with QED
\cite{Richard} that the larger the number of initial $\bar{q} q$
valence pairs which need to be annihilated for a spesific $\NbarN$
annihilation reaction to occur, the more central is the reaction.
\end{enumerate}

\medskip

Since our explanation \cite{TMK} of the maximal \Asym\
is based on a rather general picture, we expect
that the maximal \Asym\ will persist as the incident $\pbar$
energy increases.
Judging from the success of a similar diffraction model
analysis for the $\pi N \rightarrow \pi N$ scattering data for
\plab\ between 2 and 6 GeV/c \cite{Ho71},
our description is presumably valid
in a similar energy range.
However, as discussed by Carlson \etal \cite{Carl92,Myh93},
we do expect our scheme based on the hadronic picture to break down
at higher energies when the perturbative QCD regime of exclusive
hadronic reactions is reached.
The onset of the perturbative QCD regime
may be signaled by
a significant change in the energy and angular variation of the asymmetry.
Therefore the measurement of \Asym\
for \plab $>$ 2 GeV/c is expected to be extremely useful not only for
a better understanding of
the nature of the extraordinarily large asymmetry observed in the LEAR
energy region,
but also for monitoring the possible onset of perturbative QCD.
\medskip

This work is supported in part by NSF grant no. PHYS-9006844.

\hspace{1cm}



\begin{thebibliography}{99}

\bibitem{Ei75}
E. Eisenhandler \etal\ , Nucl. Phys. {\bf B96}, 109 (1975);
A. A. Carter \etal\ ,  Nucl. Phys. {\bf B127}, 202 (1977).

\bibitem{Te89}
F. Tessarotto \etal\ , Nucl. Phys. {\bf B}
(Proc. Suppl.) {\bf 8}, 141 (1989).

\bibitem{Hasan92} A. Hasan \etal\ , Nucl. Phys. {\bf B378}, 3 (1992)

\bibitem{Ca77} A. A. Carter \etal\ , Phys. Lett. {\bf B67}, 117
(1977); A. A. Carter, Phys. Lett. {\bf B67}, 122 (1977).

\bibitem{MO88} B. R. Martin and G. C. Oades,
Nucl. Phys. {\bf A483}, 669 (1988);
see also sect.4 of N. Isgur  and K. K\"onigsmann,
Nucl. Phys. {\bf A527}, 491c (1991).

\bibitem{TMK}
S. Takeuchi, F. Myhrer and K. Kubodera, AIP Conf. Proc.
{\bf 243}, 358 (1992),
4th Conf. on the
{\it Intersections Between Particle and Nuclear Physics}; and \\
Nucl. Phys. {\bf A} to appear (1993).


\bibitem{LT90}
G. Q. Liu and F. Tabakin, Phys. Rev. {\bf C41}, 665 (1990).

\bibitem{MHS91}
V. Mull, K. Holinde and J. Speth, J\"ulich preprint KFA-IKP(TH)1991-37
(1991).

\bibitem{Ho71}
H. H{\o}gaasen, Phys. Norvegica {\bf 5}, 219 (1971);
H. H{\o}gaasen and C. Michael, Nucl. Phys. {\bf B44}, 214 (1972);
V. Barger and R. J. N. Phillips, Nucl. Phys. {\bf B87}, 221 (1975).

\bibitem{Ei76}
E. Eisenhandler \etal\ ,  Nucl. Phys. {\bf B113}, 1 (1976).

\bibitem{Richard} J. Carbonell \etal\ , Zeit. Phys. {\bf A334}, 329
(1989).

\bibitem{Carl92} C. E. Carlson, M. Chachkhunashvili and F. Myhrer,
Phys. Rev. {\bf D46}, 2891 (1992).

\bibitem{Myh93} F. Myhrer, talk at the Hadron Scattering and Spin
Sessions at the Workshop on {\it Future Directions in Particle
and Nuclear Physics at Multi-GeV Hadron
Beam Facilities}, Brookhaven Nat. Lab., March (1993).

%
%
%
\end{thebibliography}
\end{document}